\documentclass[longauth]{aa}

\usepackage{txfonts}
\usepackage{graphicx}
\usepackage{array}
\usepackage{booktabs}
\usepackage{amssymb}
\usepackage{subcaption}
\usepackage{multirow}
\usepackage{url}
\usepackage[colorlinks=true,allcolors=blue]{hyperref}
\usepackage{xcolor}

\PassOptionsToPackage{hyphens}{url}

\newcommand{\ffarcs}{\mbox{\ensuremath{.\!\!^{\prime\prime}}}}

\newcolumntype{L}[1]{>{\raggedright\let\newline\\\arraybackslash\hspace{0pt}}m{#1}}
\newcolumntype{C}[1]{>{\centering\let\newline\\\arraybackslash\hspace{0pt}}m{#1}}

\begin{document}

\title{Direct imaging discovery of a young giant planet\\orbiting on Solar System scales}

\titlerunning{Direct imaging discovery of a young giant planet orbiting on Solar System scales}

\author{
 T.~Stolker\inst{\ref{leiden}}
 \and M.~Samland\inst{\ref{mpia}}
 \and L.~B.~F.~M.~Waters\inst{\ref{nijmegen}}
 \and M.~E.~van den Ancker\inst{\ref{esog}}
 \and W.~O.~Balmer\inst{\ref{jhupa},\ref{stsci}}
 \and S.~Lacour\inst{\ref{lesia},\ref{esog}}
 \and M.~L.~Sitko\inst{\ref{boulder}}
 \and J.~J.~Wang\inst{\ref{northwestern}}
 \and M.~Nowak\inst{\ref{cam}}
 \and A.-L.~Maire\inst{\ref{ipag}}
 \and J.~Kammerer\inst{\ref{esog}}
 \and G.~P.~P.~L.~Otten\inst{\ref{sinica}}
 \and R.~Abuter\inst{\ref{esog}}
 \and A.~Amorim\inst{\ref{lisboa},\ref{centra}}
 \and M.~Benisty\inst{\ref{mpia}}
 \and J.-P.~Berger\inst{\ref{ipag}}
 \and H.~Beust\inst{\ref{ipag}}
 \and S.~Blunt\inst{\ref{northwestern}}
 \and A.~Boccaletti\inst{\ref{lesia}}
 \and M.~Bonnefoy\inst{\ref{ipag}}
 \and H.~Bonnet\inst{\ref{esog}}
 \and M.~S.~Bordoni\inst{\ref{mpe}}
 \and G.~Bourdarot\inst{\ref{mpe}}
 \and W.~Brandner\inst{\ref{mpia}}
 \and F.~Cantalloube\inst{\ref{ipag}}
 \and P.~Caselli \inst{\ref{mpe}}
 \and B.~Charnay\inst{\ref{lesia}}
 \and G.~Chauvin\inst{\ref{cotedazur}}
 \and A.~Chavez\inst{\ref{northwestern}}
 \and A.~Chomez\inst{\ref{lesia},\ref{ipag}}
 \and E.~Choquet\inst{\ref{lam}}
 \and V.~Christiaens\inst{\ref{liege}}
 \and Y.~Cl\'enet\inst{\ref{lesia}}
 \and V.~Coud\'e~du~Foresto\inst{\ref{lesia}}
 \and A.~Cridland\inst{\ref{leiden}}
 \and R.~Davies\inst{\ref{mpe}}
 \and R.~Dembet\inst{\ref{lesia}}
 \and J.~Dexter\inst{\ref{boulder}}
 \and C.~Dominik\inst{\ref{amsterdam}}
 \and A.~Drescher\inst{\ref{mpe}}
 \and G.~Duvert\inst{\ref{ipag}}
 \and A.~Eckart\inst{\ref{cologne},\ref{bonn}}
 \and F.~Eisenhauer\inst{\ref{mpe}}
 \and N.~M.~F\"orster Schreiber\inst{\ref{mpe}}
 \and P.~Garcia\inst{\ref{centra},\ref{porto}}
 \and R.~Garcia~Lopez\inst{\ref{dublin},\ref{mpia}}
 \and T.~Gardner\inst{\ref{exeterAstro}}
 \and E.~Gendron\inst{\ref{lesia}}
 \and R.~Genzel\inst{\ref{mpe},\ref{ucb}}
 \and S.~Gillessen\inst{\ref{mpe}}
 \and J.~H.~Girard\inst{\ref{stsci}}
 \and S.~Grant\inst{\ref{mpe}}
 \and X.~Haubois\inst{\ref{esoc}}
 \and G.~Hei\ss el\inst{\ref{actesa},\ref{lesia}}
 \and Th.~Henning\inst{\ref{mpia}}
 \and S.~Hinkley\inst{\ref{exeter}}
 \and S.~Hippler\inst{\ref{mpia}}
 \and M.~Houll\'e\inst{\ref{cotedazur}}
 \and Z.~Hubert\inst{\ref{ipag}}
 \and L.~Jocou\inst{\ref{ipag}}
 \and M.~Keppler\inst{\ref{mpia}}
 \and P.~Kervella\inst{\ref{lesia}}
 \and L.~Kreidberg\inst{\ref{mpia}}
 \and N.~T.~Kurtovic\inst{\ref{mpe}}
 \and A.-M.~Lagrange\inst{\ref{ipag},\ref{lesia}}
 \and V.~Lapeyr\`ere\inst{\ref{lesia}}
 \and J.-B.~Le~Bouquin\inst{\ref{ipag}}
 \and D.~Lutz\inst{\ref{mpe}}
 \and F.~Mang\inst{\ref{mpe}}
 \and G.-D.~Marleau\inst{\ref{duisburg},\ref{bern},\ref{mpia}}
 \and A.~M\'erand\inst{\ref{esog}}
 \and M.~Min\inst{\ref{sron}}
 \and P.~Molli\`ere\inst{\ref{mpia}}
 \and J.~D.~Monnier\inst{\ref{umich}}
 \and C.~Mordasini\inst{\ref{bern}}
 \and D.~Mouillet\inst{\ref{ipag}}
 \and E.~Nasedkin\inst{\ref{mpia}}
 \and T.~Ott\inst{\ref{mpe}}
 \and C.~Paladini\inst{\ref{esoc}}
 \and T.~Paumard\inst{\ref{lesia}}
 \and K.~Perraut\inst{\ref{ipag}}
 \and G.~Perrin\inst{\ref{lesia}}
 \and O.~Pfuhl\inst{\ref{esog}}
 \and N.~Pourr\'e\inst{\ref{ipag}}
 \and L.~Pueyo\inst{\ref{stsci}}
 \and S.~P.~Quanz\inst{\ref{ethz}}
 \and D.~C.~Ribeiro\inst{\ref{mpe}}
 \and E.~Rickman\inst{\ref{esa}}
 \and Z.~Rustamkulov\inst{\ref{jhueps}}
 \and J.~Shangguan\inst{\ref{beijing}}
 \and T.~Shimizu \inst{\ref{mpe}}
 \and D.~Sing\inst{\ref{jhupa},\ref{jhueps}}
 \and J.~Stadler\inst{\ref{mpa},\ref{origins}}
 \and O.~Straub\inst{\ref{origins}}
 \and C.~Straubmeier\inst{\ref{cologne}}
 \and E.~Sturm\inst{\ref{mpe}}
 \and L.~J.~Tacconi\inst{\ref{mpe}}
 \and E.F.~van~Dishoeck\inst{\ref{leiden},\ref{mpe}}
 \and A.~Vigan\inst{\ref{lam}}
 \and F.~Vincent\inst{\ref{lesia}}
 \and S.~D.~von~Fellenberg\inst{\ref{bonn}}
 \and F.~Widmann\inst{\ref{mpe}}
 \and T.~O.~Winterhalder\inst{\ref{esog}}
 \and J.~Woillez\inst{\ref{esog}}
 \and S.~Yazici\inst{\ref{mpe}}
}

\institute{ 
   Leiden Observatory, Leiden University, Einsteinweg 55, 2333 CC Leiden, The Netherlands
\label{leiden}      \and
   Max Planck Institute for Astronomy, K\"onigstuhl 17, 69117 Heidelberg, Germany
\label{mpia}      \and
   Department of Astrophysics/IMAPP, Radboud University, Heyendaalseweg 135, 6525 AJ Nijmegen, The Netherlands
\label{nijmegen}      \and
   European Southern Observatory, Karl-Schwarzschild-Stra\ss e 2, 85748 Garching, Germany
\label{esog}      \and
   Department of Physics \& Astronomy, Johns Hopkins University, 3400 N. Charles Street, Baltimore, MD 21218, USA
\label{jhupa}      \and
   Space Telescope Science Institute, 3700 San Martin Drive, Baltimore, MD 21218, USA
\label{stsci}      \and
   LESIA, Observatoire de Paris, PSL, CNRS, Sorbonne Universit\'e, Universit\'e de Paris, 5 place Janssen, 92195 Meudon, France
\label{lesia}      \and
   Department of Astrophysical \& Planetary Sciences, JILA, Duane Physics Bldg., 2000 Colorado Ave, University of Colorado, Boulder, CO 80309, USA
\label{boulder}      \and
   Center for Interdisciplinary Exploration and Research in Astrophysics (CIERA) and Department of Physics and Astronomy, Northwestern University, Evanston, IL 60208, USA
\label{northwestern}      \and
   Institute of Astronomy, University of Cambridge, Madingley Road, Cambridge CB3 0HA, United Kingdom
\label{cam}      \and
   Univ. Grenoble Alpes, CNRS, IPAG, 38000 Grenoble, France
\label{ipag}      \and
   Academia Sinica, Institute of Astronomy and Astrophysics, 11F Astronomy-Mathematics Building, NTU/AS campus, No. 1, Section 4, Roosevelt Rd., Taipei 10617, Taiwan
\label{sinica}      \and
   Anton Pannekoek Institute for Astronomy, University of Amsterdam, Science Park 904, 1098 XH Amsterdam, The Netherlands
\label{amsterdam}      \and
   SRON Netherlands Institute for Space Research, Sorbonnelaan 2, 3584 CA Utrecht, The Netherlands
\label{sron}      \and
   Institute for Particle Physics and Astrophysics, ETH Zurich, Wolfgang-Pauli-Strasse 27, 8093 Zurich, Switzerland
\label{ethz}      \and
   Universidade de Lisboa - Faculdade de Ci\^encias, Campo Grande, 1749-016 Lisboa, Portugal
\label{lisboa}      \and
   CENTRA - Centro de Astrof\' isica e Gravita\c c\~ao, IST, Universidade de Lisboa, 1049-001 Lisboa, Portugal
\label{centra}      \and
   Max Planck Institute for extraterrestrial Physics, Giessenbachstra\ss e~1, 85748 Garching, Germany
\label{mpe}      \and
   Université Côte d’Azur, Observatoire de la Côte d’Azur, CNRS, Laboratoire Lagrange, Bd de l'Observatoire, CS 34229, 06304 Nice cedex 4, France
\label{cotedazur}      \and
   Aix Marseille Univ, CNRS, CNES, LAM, Marseille, France
\label{lam}      \and
  STAR Institute, Universit\'e de Li\`ege, All\'ee du Six Ao\^ut 19c, 4000 Li\`ege, Belgium
\label{liege}      \and
   1.\ Institute of Physics, University of Cologne, Z\"ulpicher Stra\ss e 77, 50937 Cologne, Germany
\label{cologne}      \and
   Max Planck Institute for Radio Astronomy, Auf dem H\"ugel 69, 53121 Bonn, Germany
\label{bonn}      \and
   Universidade do Porto, Faculdade de Engenharia, Rua Dr.~RobertoRua Dr.~Roberto Frias, 4200-465 Porto, Portugal
\label{porto}      \and
   School of Physics, University College Dublin, Belfield, Dublin 4, Ireland
\label{dublin}      \and
   Astrophysics Group, Department of Physics \& Astronomy, University of Exeter, Stocker Road, Exeter, EX4 4QL, United Kingdom
\label{exeterAstro}      \and
   Departments of Physics and Astronomy, Le Conte Hall, University of California, Berkeley, CA 94720, USA
\label{ucb}      \and
   European Southern Observatory, Casilla 19001, Santiago 19, Chile
\label{esoc}      \and
   Advanced Concepts Team, European Space Agency, TEC-SF, ESTEC, Keplerlaan 1, NL-2201, AZ Noordwijk, The Netherlands
\label{actesa}      \and
   University of Exeter, Physics Building, Stocker Road, Exeter EX4 4QL, United Kingdom
\label{exeter}      \and
   Fakult\"at f\"ur Physik, Universit\"at Duisburg-Essen, Lotharstraße 1, 47057 Duisburg, Germany
\label{duisburg}      \and
   Instit\"ut f\"ur Astronomie und Astrophysik, Universit\"at T\"ubingen, Auf der Morgenstelle 10, 72076 T\"ubingen, Germany
\label{bern}      \and
   Astronomy Department, University of Michigan, Ann Arbor, MI 48109 USA
\label{umich}      \and
   European Space Agency (ESA), ESA Office, Space Telescope Science Institute, 3700 San Martin Drive, Baltimore, MD 21218, USA
\label{esa}      \and
   Department of Earth \& Planetary Sciences, Johns Hopkins University, Baltimore, MD, USA
\label{jhueps}      \and
   Max Planck Institute for Astrophysics, Karl-Schwarzschild-Str. 1, 85741 Garching, Germany
\label{mpa}      \and
   Excellence Cluster ORIGINS, Boltzmannstraße 2, D-85748 Garching bei München, Germany
\label{origins}    \and
   The Kavli Institute for Astronomy and Astrophysics, Peking University, Beijing 100871, China
\label{beijing}
}

\date{Received ?; accepted ?}

\abstract
{HD\,135344\,AB is a young visual binary system that is best known for the protoplanetary disk around the secondary star. The circumstellar environment of the A0-type primary star, on the other hand, is already depleted. HD\,135344\,A is therefore an ideal target for the exploration of recently formed giant planets because it is not obscured by dust.}
{We searched for and characterized substellar companions to HD\,135344\,A down to separations of about 10\,au.}
{We observed HD\,135344\,A with VLT/SPHERE in the $H23$ and $K12$ bands and obtained $YJ$ and $YJH$ spectroscopy. In addition, we carried out VLTI/GRAVITY observations for the further astrometric and spectroscopic confirmation of a detected companion.}
{We discovered a close-in young giant planet, HD\,135344\,Ab, with a mass of about 10\,$M_\mathrm{J}$. The multi-epoch astrometry confirms the bound nature based on common parallax and common proper motion. This firmly rules out the scenario of a non-stationary background star. The semi-major axis of the planetary orbit is approximately 15--20\,au, and the photometry is consistent with that of a mid L-type object. The inferred atmospheric and bulk parameters further confirm the young and planetary nature of the companion.}
{HD\,135344\,Ab is one of the youngest directly imaged planets that has fully formed and orbits on Solar System scales. It is a valuable target for studying the early evolution and atmosphere of a giant planet that could have formed in the vicinity of the snowline.}


\maketitle

\section{Introduction}
\label{sec:introduction}

\begin{table*}
\caption{Observation details.}
\label{table:observations}
\centering
\bgroup
\def\arraystretch{1.25}
\begin{tabular}{L{1.9cm} C{1.6cm} C{2cm} C{2.8cm} C{1.5cm} C{1.5cm} C{1.2cm} C{0.8cm}}
\hline\hline
UT date\tablefootmark{a} & Instrument & Mode & DIT/NDIT/NEXP\tablefootmark{b} & Airmass    & Seeing\tablefootmark{c} & $\tau_0$\tablefootmark{d} & $\pi$\tablefootmark{e} \\
           &            &             &               &            & (arcsec)        & (ms)          & (deg) \\
\hline
2019 May 09 & SPHERE     & IRDIFS      & 48/3/16       & 1.02--1.03 & $0.78 \pm 0.09$ & $2.7 \pm 0.3$ & 43.8  \\
2019 Jul 06 & SPHERE     & IRDIFS      & 48/3/16       & 1.02--1.03 & $0.66 \pm 0.10$ & $5.3 \pm 0.6$ & 43.3  \\
2021 Jul 16 & SPHERE     & IRDIFS      & 4/32/25       & 1.02--1.05 & $0.64 \pm 0.11$ & $3.3 \pm 0.5$ & 78.3  \\
2022 May 04 & SPHERE     & IRDIFS\_EXT & 4/32/24       & 1.02--1.05 & $0.84 \pm 0.18$ & $4.5 \pm 1.0$ & 78.4  \\
2022 Jul 19 & GRAVITY    & ON-AXIS     & 100/4/8       & 1.03--1.13 & $1.26 \pm 0.19$ & $1.8 \pm 0.2$ & 46.0  \\
2023 May 08 & GRAVITY    & ON-AXIS     & 100/4/7       & 1.03--1.13 & $0.95 \pm 0.09$ & $3.6 \pm 0.5$ & 35.2  \\
2023 Jul 01 & GRAVITY    & ON-AXIS     & 100/4/4       & 1.08--1.17 & $0.79 \pm 0.12$ & $6.4 \pm 5.2$ & 12.2  \\
\hline
\end{tabular}
\egroup
\tablefoot{
\tablefoottext{a}{UT date at the start of the observations.}
\tablefoottext{b}{Detector integration time, number of integrations per exposure, and number of exposures. The listed values of the SPHERE observations are taken from IRDIS.}
\tablefoottext{c,d}{Sample mean and standard deviation of the DIMM seeing and coherence time.}
\tablefoottext{e}{Rotation by the parallactic angle.}
}
\end{table*}

HD\,135344\,AB is a visual binary system that is located in Upper Centaurus Lupus (UCL) region of the Sco-Cen OB association. The secondary F4-type star, HD\,135344\,B, has been studied for several decades because of its prominent IR excess. During more recent years, the protoplanetary disk was spatially resolved and revealed a central cavity \citep{brown2009,garufi2013}, spiral arms \citep{muto2012}, and variable shadowing by the inner disk \citep{stolker2017b}. These disk features might indicate planet-disk interactions, but the suspected planets have remained hidden \citep[e.g.,][]{maire2017,cugno2024}.

While planet formation appears to be ongoing at HD\,135344\,B, the circumstellar environment of the A0-type primary star, HD\,135344\,A, is already largely depleted, given the absence of strong IR excess in the spectral energy distribution (SED). HD\,135344\,A and~B are proper-motion binary partners \citep{mason2001} with an angular separation of 21\ffarcs2 ($\approx$2800~au), and their circumstellar disks have therefore likely evolved independently, depending on the eccentricity of the orbits. We note that the secondary star has incorrectly been referred to as HD\,135344 in some cases, although the issue had already been pointed out by \citet{coulson1995}.

The two stars in the HD\,135344\,AB binary system are expected to be coeval. The pre-main-sequence age of the secondary star can therefore be adopted as the age of the main-sequence primary star. \citet{garufi2018} inferred an age of $11.9_{-5.8}^{+3.7}$~Myr for HD\,135344\,B by using stellar evolutionary tracks and the Gaia parallax. In addition to the young age, giant planets on wide orbits are most commonly detected around intermediate-mass stars \citep[e.g.,][]{nielsen2019,vigan2021}. The youth, spectral type, and dust-depleted environment make HD\,135344\,A an excellent target to search for young giant planets.

We report high-contrast imaging observations with which we explore the circumstellar environment of HD\,135344\,A for the first time. We discovered a young giant planet that we confirmed through a detailed astrometric and spectral analysis. The evidence for the bound and planetary nature accumulated based on the observations and results. For simplicity, we refer to the discovered source as HD\,135344\,Ab from here on.

\section{Observations and data reduction}
\label{sec:observations}

\subsection{VLT/SPHERE high-contrast imaging}
\label{sec:sphere_observations}

HD\,135344\,A was observed with VLT/SPHERE \citep{beuzit2019} on the nights of 2019 May 8, 2019 July 5, 2021 July 16, and 2022 May 3. The first observations were carried out with the \texttt{IRDIFS} mode so that we could benefit from the highest angular resolution. We used the IRDIS dual-band camera with the the $H23$ filters \citep{dohlen2008,vigan2010}, and we obtained a low-resolution $YJ$ spectrum ($R \approx 50$) with IFS \citep{claudi2008}. The first observation, which led to the discovery, did not fully meet the requested conditions (and hence, it was repeated in July), but the quality was sufficient for a robust detection. We then repeated the observation in 2021 to confirm the source at the same wavelength. In 2022, we used the \texttt{IRDIFS\_EXT} mode to obtain $K12$ dual-band imaging and a low-resolution $YJH$ spectrum ($R \approx 30$).

The integration times and observing conditions are listed in Table~\ref{table:observations}. We used deep exposures with a detector integration time (DIT) of 48 and 64 seconds for the simultaneous IRDIS and IFS measurements, respectively, in 2019. After we detected the source at a small separation, we decided to change the observing strategy. Specifically, we used a DIT of 4 and 6 seconds for IRDIS and IFS, respectively, to sample the speckle variation on a faster timescale, which is beneficial for the post-processing. We also left the satellite spots on throughout the observations to enable a more accurate centering and flux calibration.

The IRDIS and IFS data were reduced with \texttt{vlt-sphere}\footnote{\url{https://github.com/avigan/SPHERE}} \citep{vigan2020}, which provides a Python wrapper for the \texttt{EsoRex} recipes. It also applies a recalibration of the IFS wavelength solution, and it centers the coronagraphic frames based on the satellite spots. We post-processed the IRDIS data with \texttt{PynPoint}\footnote{\url{https://github.com/PynPoint/PynPoint}} by applying full-frame principal component analysis (PCA) to subtract the stellar halo and speckles \citep{amara2012,stolker2019}. For the astrometric and photometric measurements, we followed the procedure outlined by \citet{stolker2020a}, which corrects for self-subtraction and includes the systematic uncertainty in the error budget. The extracted fluxes were corrected for the coronagraph throughput, which ranged from 97\% to 93\% as the separation of the planet decreased.

The IFS data were post-processed with \texttt{TRAP}\footnote{\url{https://github.com/m-samland/trap}}, which specifically is a more powerful detection and calibration technique for sources at small separation because systematics are modeled in the temporal instead of the spatial domain \citep{samland2021}. We analyzed the $YJ$ and $YJH$ spectra from 2021 and 2022, respectively, because these datasets were obtained with continuous satellite spots. Similar to the IRDIS data, the amplitudes of the spots were used to identify low-quality frames and their temporal variation was accounted for in the spectral extraction. The wavelength-averaged S/N of the discovered source is 2.9 and 1.5 for the $YJ$ and $YJH$ spectra, respectively. The higher S/N of the $YJ$ spectrum might be due to the better seeing conditions.

The contrasts were converted into magnitudes and fluxes by using a synthetic stellar spectrum and a flux-calibrated spectrum of Vega \citep{bohlin2014}. Table~\ref{table:stellar_param} in Appendix~\ref{app:stellar_parameters} lists the inferred stellar parameters and the synthetic magnitudes of HD\,135344\,A in the IRDIS filters that were computed from the posterior distributions. The synthetic IFS spectrum of HD\,135344\,A was extracted by sampling random spectra from the posterior, smoothing them to $R = 30$, and rebinning to the wavelength solution of the instrument.

\begin{figure}
\centering
\includegraphics[width=\linewidth]{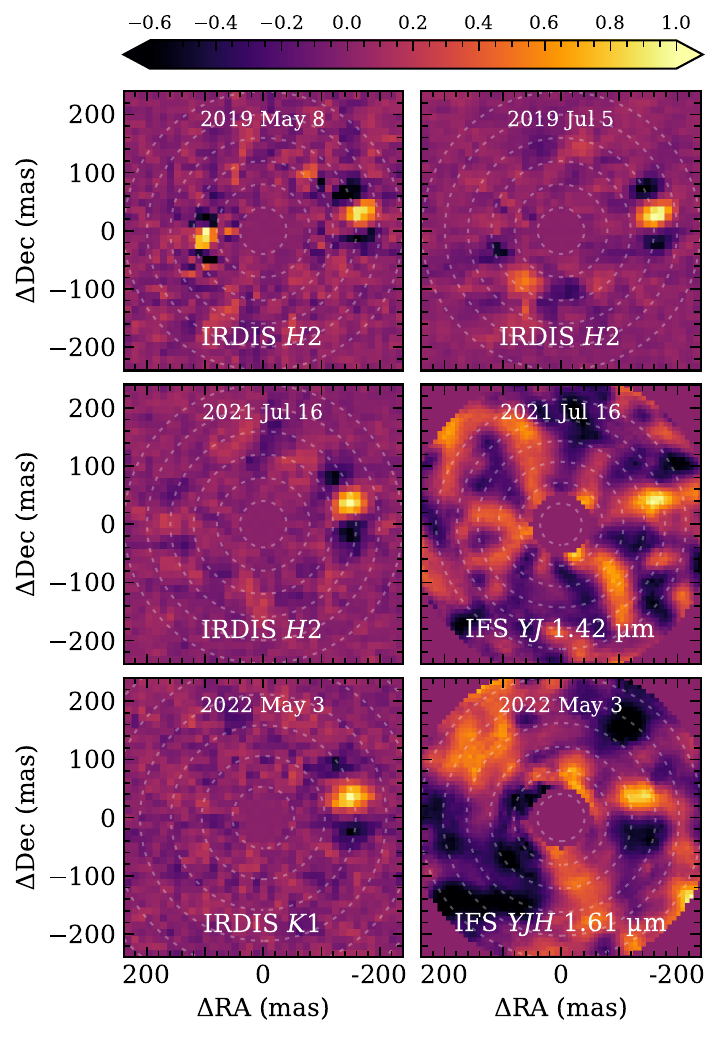}
\caption{Detections of HD\,135344\,Ab with VLT/SPHERE. For IRDIS, the images show the residuals from the PSF subtraction for one of the dual-band filters. For IFS, the images show the detection maps for one of the wavelength channels. The planet is seen in westward direction (i.e., toward the right). The color scale is linear and normalized to the brightest pixel in each image. The dotted circles indicate the separation from the central star in integer multiples of $\lambda/D$. The night of the observation is given in each panel.}
\label{fig:sphere}
\end{figure}

\begin{figure*}
\centering
\includegraphics[width=\linewidth]{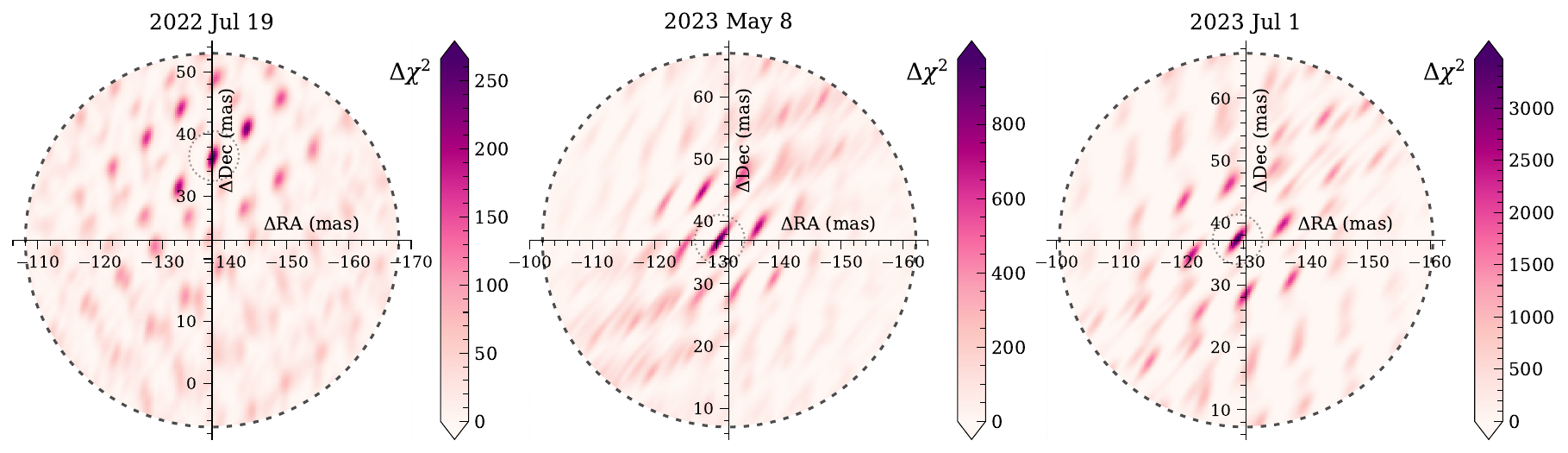}
\caption{Detections of HD\,135344\,AB with VLTI/GRAVITY. The color scale of the detection maps shows $\Delta\chi^2 = \chi_\mathrm{no\,planet} - \chi_\mathrm{planet}$. The intersection of the axes is at the center of the fiber, which has a field of view of $\approx$60~mas. For each observation, the feature with the highest likelihood is encircled, which corresponds to the planet position. The night of the observations is shown at the top of each panel.}
\label{fig:gravity}
\end{figure*}

\subsection{VLTI/GRAVITY dual-field interferometry}
\label{sec:gravity_observations}

The system was also observed with the four unit telescopes (UTs) of VLTI/GRAVITY on the nights of 2022 July 19, 2023 May 8, and 2023 July 1 (see Table~\ref{table:observations}). GRAVITY provides exquisite astrometric precision for directly imaged planets and medium-resolution ($R \approx 500$) $K$-band spectroscopy \citep{gravity2017}. We used the dual-field on-axis mode with the fringe-tracking fiber centered on the star and the science fiber alternating between the star and planet. This strategy enables referencing and accurate calibration of the interferometric visibilities \citep{gravity2020,lacour2020}. For the third observation, we benefitted from the recently commissioned faint mode, which turns the metrology lasers off during the science exposures. This yields a detection with a higher S/N of faint companions \citep{widmann2022}.

Standard data reduction procedures were applied with the \texttt{run{\textunderscore}gravi{\textunderscore}reduce} script\footnote{\url{https://www.eso.org/sci/facilities/paranal/instruments/gravity/tools.html}}
that calls the \texttt{EsoRex} recipes. We then used the \texttt{exoGravity}\footnote{\url{https://gitlab.obspm.fr/mnowak/exogravity.git}} pipeline for phase referencing, subtracting the stellar component from the visibilities, and extracting the astrometry and spectrum \citep{gravity2019,gravity2020,nowak2020}. After calibrating the full datasets, we excluded the baselines of UT3 from the 2023 May data because of an issue with the metrology. The wavelength-averaged S/N of the spectra of the discovered source is 1.1, 2.6, and 5.0 (in chronological order). We only included the GRAVITY spectrum from 2023 July in the analysis because a covariance-weighted combination with the other two spectra did not improve the S/N, in particular, because the systematics in these two spectra are stronger. Similar to the IFS flux calibration, we used a synthetic spectrum to convert the contrast spectrum into fluxes while taking the uncertainties on the stellar spectrum into account.

\section{Results}
\label{sec:results}

\subsection{Direct detection of a young giant planet}
\label{sec:planet_detection}

\begin{table*}
\caption{Astrometry of HD\,135344\,Ab.}
\label{table:astrometry}
\centering
\bgroup
\def\arraystretch{1.25}
\begin{tabular}{L{1.9cm} C{1.6cm} C{2.6cm} C{2.2cm} C{2.2cm} C{1cm}}
\hline\hline
UT date\tablefootmark{a} & MJD & Instrument & $\Delta$RA\tablefootmark{b} & $\Delta$Dec\tablefootmark{c} & $\rho$\tablefootmark{d} \\
 & & & (mas) & (mas) & \\
\hline
2019 May 09 & 58612.2 & SPHERE/IRDIS & $-161.6 \pm 2.7$ & $31.6 \pm 1.3$ & $-0.33$ \\
2019 Jul 06 & 58670.0 & SPHERE/IRDIS & $-162.0 \pm 2.0$ & $28.2 \pm 1.3$ & $-0.17$ \\
2021 Jul 16 & 59412.0 & SPHERE/IRDIS & $-144.8 \pm 1.9$ & $36.1 \pm 1.2$ & $-0.22$ \\
2022 May 04 & 59703.2 & SPHERE/IRDIS & $-142.3 \pm 5.2$ & $35.8 \pm 2.0$ & $-0.58$ \\
2022 Jul 19 & 59779.06 & GRAVITY & $-138.21 \pm 0.15$ & $36.34 \pm 0.17$ & $-0.58$ \\
2023 May 08 & 60072.15 & GRAVITY & $-130.41 \pm 0.20$ & $37.00 \pm 0.28$ & $-0.95$ \\
2023 Jul 01 & 60126.98 & GRAVITY & $-129.07 \pm 0.13$ & $37.36 \pm 0.08$ & $-0.97$ \\
\hline
\end{tabular}
\egroup
\tablefoot{
\tablefoottext{a}{UT date at the start of the observations.}
\tablefoottext{b,c}{Coordinates, RA and Dec, relative to the star HD\,135344\,A.}
\tablefoottext{d}{The Pearson correlation coefficient, $\rho$, quantifies the correlation between the uncertainties of $\Delta$RA and $\Delta$Dec.}
}
\end{table*}

The residuals of the PSF subtraction from the imaging observations with SPHERE are shown in Fig.~\ref{fig:sphere}. In 2019, we detected an off-axis point source, HD\,135344\,Ab, which is located west of the star at an approximate separation of $4\lambda/D$ in $H23$. In 2021, the source was detected again, but at a somewhat smaller separation, and in 2022, it was also detected with the $K12$ filters. The PSF shape is typical for a post-processed point source. The negative lobes are due to the inherent self-subtraction effect by angular differential imaging (ADI). The source clearly stands out against the fainter speckle field in the background and is detected with $\mathrm{S/N} \approx 10$ in all IRDIS imaging data. We did not detect extended emission in the data, indicating that the circumstellar environment is indeed depleted in small dust. A feature is also visible in the IRDIS image from 2019 May in the eastward direction, at a separation of $2.5\lambda/D$ from the star. It appears to be noisier than HD\,135344\,Ab and self-subtracts more quickly with an increasing number of components. Two months later, it was no longer detected, suggesting it was likely a speckle residual, as neither a background star nor a planet could have moved enough to become obscured by the coronagraph.

The SPHERE data were not fully conclusive for a confirmation that the discovered source was bound to HD\,135344\,A. We therefore decided to follow it up with the GRAVITY instrument. Figure~\ref{fig:gravity} shows the detection maps of the GRAVITY observations. The planet is clearly detected in all datasets. We adopted the feature with the highest $\Delta\chi^2$ as the position of the planet, and the contrast spectrum was also extracted at this position. The pattern and elongated shapes in the detection maps depend on the coverage of the $(u,v)$ plane (i.e., the UT baselines and field rotation). For the observation in 2022, the pointing of the fiber was 13.5~mas south of the actual location of the planet, which resulted in a fiber coupling efficiency of 0.89. For the other two observations, the fiber position was (almost) spot-on with the planet due to the improved orbital constraint.

\subsection{Relative astrometry: Planet or background star?}
\label{sec:relative astrometry}

\begin{figure}
\centering
\includegraphics[width=\linewidth]{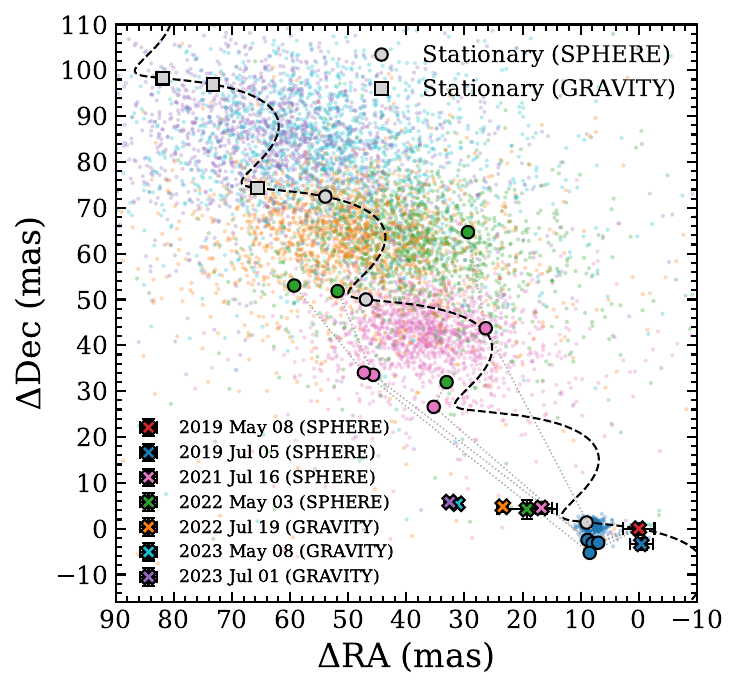}
\caption{Astrometric measurements relative to the first epoch. The crosses show the positions of HD\,135344\,Ab, which moves eastward. The colored circles show the positions of the suspected background sources in the IRDIS field of view, which are connected with dotted lines between epochs. The dashed line shows the track for a stationary background source, and the gray circles and squares indicate the three SPHERE and three GRAVITY epochs, respectively, after the initial detection. The small dots represent a sample of Gaia sources within 0.1 deg from HD\,135344\,A (see main text for details). The colors indicate a specific epoch, for example, all pink markers correspond to 2021 Jul 16.}
\label{fig:relative_astrometry}
\end{figure}

HD\,135344\,AB is located in the Sco-Cen OB association, which is close to the direction of the Galactic center. There are typically several, if not many, background sources in the 12\arcsec\,$\times$ 12\arcsec\,field of view of the IRDIS camera. In addition to the inner source identified in Sect.~\ref{sec:planet_detection}, we detected four more sources. We extracted their astrometry by directly fitting a 2D Gaussian model to the sources in the derotated frames.

The astrometric measurements of HD\,135344\,Ab are listed in Table~\ref{table:astrometry} and are shown in Fig.~\ref{fig:relative_astrometry} together with the suspected background sources. This figure displays the astrometry relative to the first epoch and shows that the planet moves mostly eastward, whereas the background sources move northeast. The direction of the background sources matches the stationary track, but it shows that the proper motions of the background sources are not negligible. This is also seen in the sample of Gaia sources, from which we adopted the proper motions to create a sample of non-stationary background stars.

From this comparison, we conclude that the relative astrometry of the IRDIS background sources is approximately consistent with the population of Gaia sources. We did not take differences in the parallax into account, which we expect to be the reason that the IRDIS background sources do not match the bulk of the Gaia sources exactly. There might also be systematics in the astrometry, such as the true north uncertainty, which affects sources far from the on-axis science target more strongly.

The planet appears to move distinctly from the background sources, although somewhat in the same direction. This initially seemed suspicious. For the last epoch, the simulated positions of the bulk of the Gaia sources differ by more than 100~mas from the position of the planet. The sample includes however a few outliers for which the magnitude and direction of the proper motions are similar to HD\,135344\,A. This yields a probability of $\approx$0.1\% that the putative planet instead is a background star with an unusually high proper motion. It is important to consider and rule out a peculiar proper motion scenario, as was shown by \citet{nielsen2017} for the case of HD\,131399\,A.

\subsection{Ruling out a non-stationary background star}
\label{sec:background_fit}

The analysis in Sect.~\ref{sec:relative astrometry} showed that the astrometric measurements are difficult to explain with a background star. In this section, we statistically explore the scenario of a background (or foreground) object by fitting the relative astrometry of HD\,135344\,Ab with a non-stationary background model. We used \texttt{backtracks}\footnote{\url{https://github.com/wbalmer/backtracks}} \citep{backtracks_zenodo} to determine the coordinates, proper motion, and parallax that best describe the astrometry as a background object. The code uses dynamic nested sampling with the \texttt{dynesty} package \citep{speagle2020} to estimate the parameter posteriors. At the start, the Gaia catalog is queried for the parallaxes and proper motions of sources within a window of 0.2~deg centered on HD\,135344\,A, and these are used as priors with the Bayesian inference, $\varpi_\mathrm{Gaia} = 0.4 \pm 0.8$~mas, $\mu_\mathrm{RA,Gaia} = -5.1 \pm 5.6$~mas~yr$^{-1}$, and $\mu_\mathrm{Dec,Gaia} = -3.8 \pm 4.2$~mas~yr$^{-1}$. We note that a similar sample was used for the proper motions applied in Fig.~\ref{fig:relative_astrometry}.

\begin{figure}
\centering
\includegraphics[width=\linewidth]{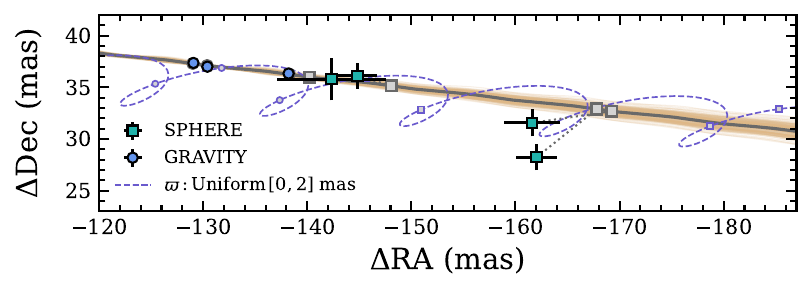}
\caption{Background fit of the relative astrometry. The figure shows 200 randomly drawn background tracks from the posterior. The track calculated from the median parameters is shown as the solid gray line. The astrometric measurements are shown with colored markers, and their respective epochs of the best-fit model are shown with gray markers. For comparison, the dashed blue line shows the best-fit model for a parallax prior that forces the source to the background ($\geq$500\,pc).}
\label{fig:ra_dec_background}
\end{figure}

Figure~\ref{fig:ra_dec_background} shows a random selection of posterior background tracks in comparison with the astrometry. The tracks are close to linear because the inferred parallax, $\varpi = 7.5 \pm 0.1$~mas, of the modeled background source is consistent with the parallax of HD\,135344\,Ab, $\varpi_\ast = 7.41 \pm 0.04$~mas. This result is in particular driven by the linear displacement of the GRAVITY astrometry, which was obtained about two months apart. Specifically, the high-precision measurements rule out a helix shape due to a heliocentric parallax of a background star. This is the first confirmation of a directly imaged planet by a common parallax to our knowledge.

The retrieved proper motion, $\mu_\mathrm{RA} = -9.0 \pm 0.1$ and $\mu_\mathrm{Dec} = -22.9 \pm 0.1$~mas~yr$^{-1}$, further confirms that the object is comoving with HD\,135344\,A. While $\mu_\mathrm{RA}$ is consistent with the prior distribution from Gaia, which is also shown in Fig.~\ref{fig:relative_astrometry}, $\mu_\mathrm{Dec}$, on the other hand, is a 4.5$\sigma$ outlier with respect to the Gaia sources. Therefore, the analysis shows that the object is most consistent with the proper motion of HD\,135344\,A. The difference is attributed to the orbital motion, which is approximately linear in the eastward direction. The background fit used the proper motion parameter to mimic the orbital movement, which caused the difference in particular to the RA component of the stellar proper motion, $\mu_\mathrm{RA,\ast} = -18.74 \pm 0.05$~mas~yr$^{-1}$ and $\mu_\mathrm{Dec,\ast} = -24.01 \pm 0.04$~mas~yr$^{-1}$ \citep{gaiadr3}.

To further quantify the significance, we list in Table\,\ref{table:evidence} the Bayesian evidence (i.e., the marginalized likelihood), $\ln \mathcal{Z}$, for three cases of the background fit. For the stationary model, we fixed the parallax and proper motion to zero. The parameter estimation informed by the Gaia priors is the fit that we described earlier in this section. For comparison, loosening the priors to uniform distributions increases $\ln \mathcal{Z}$ because the measurements did not match the bulk of the Gaia sample. Instead, as we describe in the next section in more detail, the Bayes factor of the model comparison between the orbit and background fit is $\Delta \ln \mathcal{Z} \geq 30$. We therefore conclude that the evidence is strong that the source is comoving with and orbiting HD\,135344\,A.

% The important take-away is that the tracks are close to linear because the best-fit off-axis source has a common parallax with HD\,135344\,A, while the proper motion parameter mimics the approximate linear orbital movement.

\begin{table}
\caption{Bayesian evidence}
\label{table:evidence}
\centering
\bgroup
\def\arraystretch{1.25}
\begin{tabular}{L{4.0cm} C{2.5cm}}
\hline\hline
Model & $\ln \mathcal{Z}$ \\
\hline
Keplerian orbit & $-34.6 \pm 0.1$ \\
Background: uniform priors & $-64.6 \pm 0.2$ \\
Background: Gaia priors & $-81.9 \pm 0.2$ \\
Background: stationary & $-272642.7 \pm 0.1$ \\
\hline
\end{tabular}
\egroup
\end{table}

\subsection{Orbital analysis}
\label{sec:orbital_analysis}

The discovered planet shows clear orbital movement, possibly even with a slight curvature. We therefore carried out a fit with \texttt{orbitize!}\footnote{\url{https://github.com/sblunt/orbitize}} \citep{blunt2020} to infer its orbital elements. The posterior distributions were sampled with the nested sampling algorithm from \texttt{MultiNest} \citep{feroz2008,buchner2014} while marginalizing over the parallax and system mass (see Table~\ref{table:stellar_param}). We used 2000 live points and restricted the priors on the argument of periastron, $\omega$, and longitude of the ascending node, $\Omega$, to one of the two solutions.

\begin{figure}
\centering
\includegraphics[width=\linewidth]{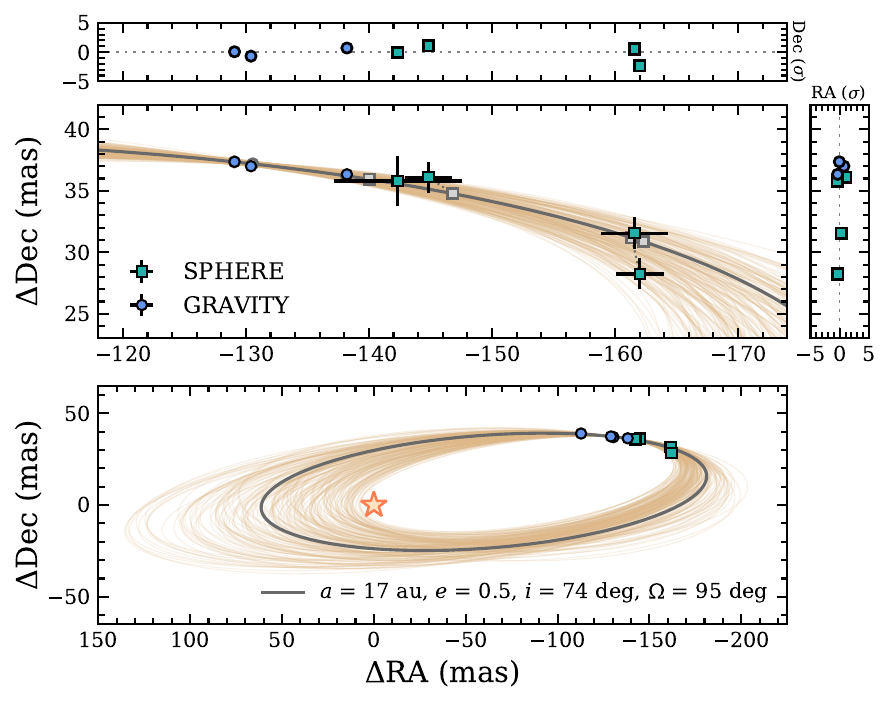}
\caption{Orbit fit of the relative astrometry. The bottom panel shows the full orbits, and the top panel shows a zoom of the observation epochs. Both panels show the same 200 orbit samples that were randomly drawn from the posterior. The orbit with the highest likelihood is shown as the solid gray line, and the residuals of the best fit are normalized by the data uncertainties. The astrometric measurements are shown with colored markers, and their respective epochs of the best-fit model are shown with gray markers and are connected with dotted lines. The planet moves in counterclockwise direction.}
\label{fig:ra_dec_orbit}
\end{figure}

Figure~\ref{fig:ra_dec_orbit} shows the posterior orbits, and the credible regions are provide in Table\,\ref{table:planet_param}. We retrieved a semi-major axis of $a = 16.5_{-2.0}^{+2.8}$~au, which corresponds to a period of $P = 45_{-8}^{+11}$~yr. The eccentricity is hardly constrained and is negatively correlated with the inclination, which is a common outcome of fitting astrometry that covers a small fraction of the orbit \citep{ferrer-chavez2021}. This is also shown in Fig.~\ref{fig:ra_dec_orbit}: Circular orbits are symmetric with respect to the star, whereas eccentric orbits have a smaller periastron. The orientation of the orbit on the sky is set by the inclination, $i \approx 74$~deg, and the longitude of the ascending node, $\Omega \approx 95$~deg. Since we only fit relative astrometry, there is a second solution for $\Omega$ and $\omega$ with a difference of $\approx$180~deg.

\subsection{Photometric analysis}
\label{sec:photometric_analysis}

\begin{table*}
\caption{Photometry of HD\,135344\,Ab.}
\label{table:photometry}
\centering
\bgroup
\def\arraystretch{1.25}
\begin{tabular}{L{1.9cm} C{1.5cm} C{1.8cm} C{2.3cm} C{2.8cm}}
\hline\hline
UT date & Filter & Contrast & App. magnitude & Flux\\
 & & (mag) & (mag) & (W\,m$^{-2}$~$\mu$m$^{-1}$)\\
\hline
\multirow{2}{*}{2019 May 09} & IRDIS $H2$ & $10.17 \pm 0.10$ & $17.76 \pm 0.11$ & $1.02 \pm 0.10$ $\times$ $10^{-16}$ \\
 & IRDIS $H3$ & $10.05 \pm 0.12$ & $17.63 \pm 0.12$ & $9.66 \pm 1.07$ $\times$ $10^{-17}$ \\
\multirow{2}{*}{2019 Jul 06} & IRDIS $H2$ & $9.95 \pm 0.11$ & $17.54 \pm 0.11$ & $1.24 \pm 0.13$ $\times$ $10^{-16}$ \\
 & IRDIS $H3$ & $9.83 \pm 0.08$ & $17.41 \pm 0.08$ & $1.18 \pm 0.09$ $\times$ $10^{-16}$ \\
\multirow{2}{*}{2021 Jul 16} & IRDIS $H2$ & $10.01 \pm 0.13$ & $17.60 \pm 0.14$ & $1.17 \pm 0.15$ $\times$ $10^{-16}$ \\
 & IRDIS $H3$ & $9.79 \pm 0.09$ & $17.37 \pm 0.09$ & $1.23 \pm 0.10$ $\times$ $10^{-16}$ \\
\multirow{2}{*}{2022 May 04} & IRDIS $K1$ & $9.30 \pm 0.10$ & $16.87 \pm 0.10$ & $8.49 \pm 0.81$ $\times$ $10^{-17}$ \\
 & IRDIS $K2$ & $8.89 \pm 0.13$ & $16.45 \pm 0.13$ & $9.65 \pm 1.16$ $\times$ $10^{-17}$ \\
 \hline
\end{tabular}
\egroup
\end{table*}

\begin{figure}
\centering
\includegraphics[width=\linewidth]{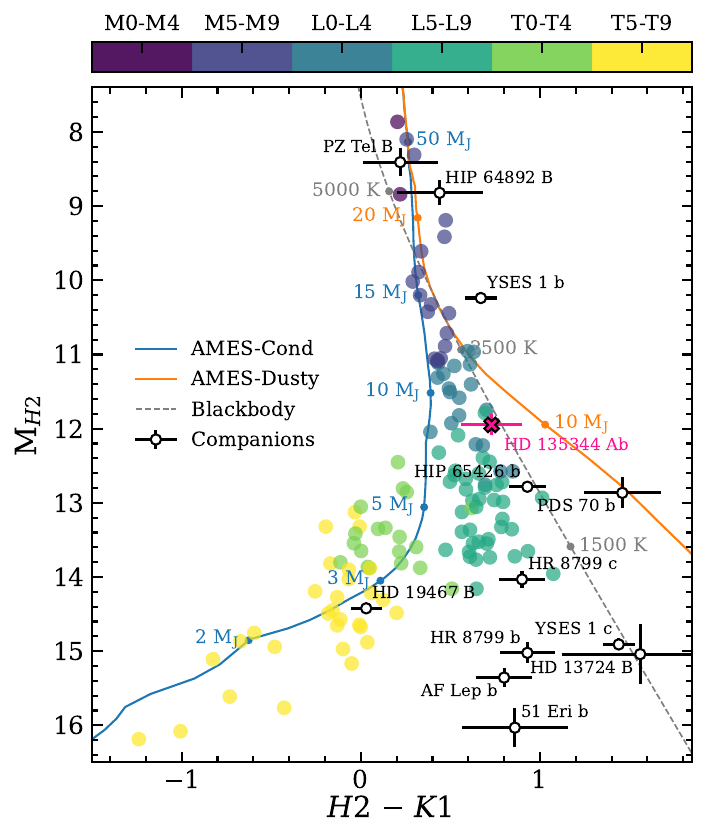}
\caption{Color--magnitude diagram of M$_{H2}$ vs. $H2$~--~$K1$. The field objects are color-coded by M, L, and T spectral types, and the directly imaged companions are labeled individually. HD\,135344\,Ab is highlighted with a pink cross. The blue and orange lines show the synthetic colors computed from the AMES-Cond and AMES-Dusty evolutionary tracks at an age of 12~Myr. Blackbody emission is shown for an object with a radius of 1~$R_\mathrm{J}$ (dashed gray line).}
\label{fig:color_magnitude}
\end{figure}

The contrast measurements and calibrated photometry are listed in Table~\ref{table:photometry} as magnitudes and fluxes. The dual-band imaging with SPHERE in the $H23$ and $K12$ bands enabled a photometric characterization of the planet, for which we adopted the magnitudes from 2021 and 2022 because these data were obtained with a dedicated strategy for an optimized accuracy of the flux calibration (see Sect.~\ref{sec:sphere_observations}). Figure~\ref{fig:color_magnitude} shows a color--magnitude diagram that was created with the \texttt{species}\footnote{\url{https://github.com/tomasstolker/species}} toolkit (see \citealt{stolker2020a} for details). In the figure, HD\,135344\,Ab is compared with late-type field objects, other directly imaged companions, and synthetic photometry from models. The isochrones that were used to calculate the synthetic fluxes were interpolated from the AMES-Cond and AMES-Dusty grids. These are evolutionary models with a cloudless and a cloudy atmosphere, respectively, as the boundary condition for the interior structure \citep{chabrier2000,allard2001,baraffe2003}.

The absolute flux and color of HD\,13534\,Ab are consistent with the field objects that have a mid-L spectral type. The photospheric temperature of L-type giant planets and brown dwarfs allows for the condensation of refractory species, and their dusty atmospheres therefore cause a red photometric appearance compared to cloudless atmospheres. The atmospheric reddening by clouds is typically stronger for young objects because their surface gravity is lower (e.g., HIP~65426~b; \citealt{chauvin2017}). The $H2$--$K1$ color of HD\,135344\.Ab is not unusually red and consistent with the field objects, although it is in particular similar to the reddest objects of that sample. At an age of 12~Myr, the $H2$ luminosity of HD\,135344\,Ab is consistent with a planetary-mass object of $\approx$10~$M_\mathrm{J}$. We provide a statistical inference of the planetary mass in Sect.~\ref{sec:evolution_bulk}.

\begin{figure*}
\centering
\includegraphics[width=\linewidth]{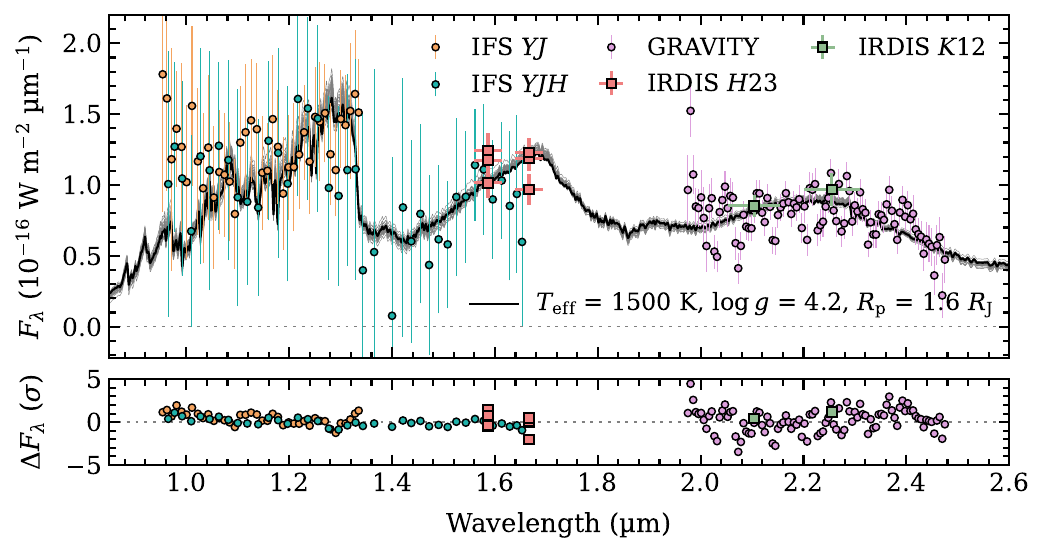}
\caption{Near-infrared spectral energy distribution of HD\,135344\,Ab. The black line shows the best-fit model spectrum from Sonora Diamondback, and the gray lines show 30 random samples from the posterior distribution, both shown at $R = 500$. The SPHERE/IFS and GRAVITY spectra are shown with circular markers, and the latter are downsampled for clarity. The SPHERE/IRDIS photometry is shown by square markers, with horizontal error bars indicating the FWHM of the corresponding filter profiles. The bottom panel shows the residuals of the best-fit model, calculated at the resolution and wavelength sampling of the data.}
\label{fig:sed_planet}
\end{figure*}

\begin{table}
\caption{Planet parameters of HD\,135344\,Ab.}
\label{table:planet_param}
\centering
\bgroup
\def\arraystretch{1.25}
\begin{tabular}{L{2.0cm} C{2.2cm} C{1.5cm}}
\hline\hline
Parameter & Value & Units \\
\hline
\multicolumn{3}{l}{\emph{Orbit fit}} \\
$a$ & $16.5_{-2.0}^{+2.8}$ & au \\
$e$ & $0.5_{-0.2}^{+0.2}$ &  \\
$i$ & $73.6_{-4.7}^{+2.8}$ & deg \\
$\Omega$\tablefootmark{a} & $94.9_{-3.1}^{+1.8}$ & deg \\
$\omega$\tablefootmark{b} & $8.2_{-10.8}^{+8.7}$ & deg \\
$t_\mathrm{p}$ & $65405_{-1378}^{+2026}$ & MJD \\
\hline
\multicolumn{3}{l}{\emph{Atmosphere fit}} \\
$T_\mathrm{eff}$ & $1510_{-35}^{+35}$ & K \\
$\log{g}$ & $4.14_{-0.07}^{+0.06}$ & dex \\
$\mathrm{[M/H]}$ & $\gtrsim$0.0 & dex \\
$f_\mathrm{sed}$ & $\lesssim$1.5 & \\
$R_\mathrm{p}$ & $1.60_{-0.06}^{+0.07}$ & $R_\mathrm{J}$ \\
$A_V$ & $\lesssim$0.5 & mag \\
\hline
\multicolumn{3}{l}{\emph{Evolution fit}\tablefootmark{c}} \\
Age & $12_{-4}^{+3}$ & Myr \\
$M_\mathrm{p}$ & $10.0_{-1.9}^{+1.4}$ & $M_\mathrm{J}$ \\
$T_\mathrm{eff}$ & $1585_{-73}^{+82}$ & K \\
$\log{g}$ & $4.1_{-0.1}^{+0.1}$ & dex \\
$R_\mathrm{p}$ & $1.45_{-0.03}^{+0.06}$ & $R_\mathrm{J}$ \\
\hline
\end{tabular}
\egroup
\tablefoot{The listed values are the median and the 16th and 84th percentiles from the posterior distributions. We only accounted for the statistical uncertainties. The lower and upper limits are provided as the 16th and 84th percentile, respectively.
\tablefoottext{a,b}{The table includes one solution for $\omega$ and $\Omega$, but there is a second solution with an offset of 180~deg.}
\tablefoottext{c}{The age and planet mass were free parameters, whereas $T_\mathrm{eff}$, $R$, and $\log\,g$ were interpolated from the evolutionary grid based on the posterior samples.}
}
\end{table}

\subsection{Atmospheric modeling}
\label{sec:atmospheric_modeling}

The spectral appearance and inferred atmospheric parameters provide further insight into the nature of HD\,135344\,Ab. We compiled the near-infrared SED in Fig.~\ref{fig:sed_planet} by combining the SPHERE and GRAVITY data. We used the Bayesian framework of the \texttt{species} toolkit \citep{stolker2020a} to fit the data with an atmospheric model, specifically, by interpolating a grid of synthetic spectra from Sonora Diamondback \citep{morley2024}. This is a radiative-convective equilibrium model that accounts for the condensation of refractory species into cloud particles. The vertical density profile of the cloud deck is parameterized by the sedimentation efficiency, $f_\mathrm{sed}$. The model uses chemical equilibrium, which is a reasonable assumption in the temperature regime of HD\,135344\,Ab, where CO will be the dominant carbon-bearing species. The parameters were estimated with the nested sampling algorithm from \texttt{MultiNest} \citep{feroz2008,buchner2014}, using 2000 live points and accounting for the spectral covariances.

The photometry and spectra are compared with the best-fit model spectrum in Fig.~\ref{fig:sed_planet}, which has a goodness-of-fit statistic of $\chi^2_\nu = 1.3$. The IFS spectra have a low S/N (see Sect.~\ref{sec:sphere_observations}), but the broad H$_2$O absorption feature between the $J$ and $H$ bands is visible in the $YJH$ spectrum. The GRAVITY spectrum also shows slopes in the pseudo-continuum that are expected to be caused by H$_2$O opacities. The CO bandheads might tentatively be detected in the $K$ band, but the GRAVITY spectrum also shows correlated noise that appears with a frequency and amplitude that might mimic the CO bands. The initial residuals indeed showed that the systematics were not fully accounted for by the covariances, and we therefore fit an uncertainty inflation for the GRAVITY spectrum. This yielded a $10 \pm 2$\% increase relative to the model fluxes. From the photometry, the $H3$ flux of the first epoch is in particular discrepant with the best-fit model, possibly due to the poorer observing conditions and because continuous satellite spots for the calibration were lacking (see Sect.~\ref{sec:sphere_observations}).

The inferred parameters are listed in Table~\ref{table:planet_param}. The temperature, $T_\mathrm{eff} \approx 1510$~K, is consistent with a mid L-type object, as empirically estimated from the $H2$ brightness in Fig.~\ref{fig:color_magnitude}. The surface gravity, $\log\,g$, and metallicity, $\mathrm{[M/H]}$, are challenging to constrain from the low-resolution spectra because this requires a highly accurate calibration, while the parameters might easily be biased otherwise. We therefore adopted the constraint from the evolution fit, $\log\,g = 4.1 \pm 0.1$ (see Sect.~\ref{sec:evolution_bulk}), as the normal prior for the atmospheric fit. The likelihood of the metallicity peaked at the supersolar edge of the model grid, but the data are also consistent with the solar abundances. The low sedimentation parameter clearly favors a dusty atmosphere, while we can rule out an atmosphere with strongly settled clouds. When we repeated the fit with a fixed sedimentation parameter of $f_\mathrm{sed} = 8$, this resulted in a Bayes factor of $\Delta \ln \mathcal{Z} = 43$ relative to the model in which $f_\mathrm{sed}$ was a free parameter.

From the $T_\mathrm{eff}$ and $R$ posterior, we computed the bolometric luminosity, $\log\,L/L_\odot = -3.9 \pm 0.1$. We note that the statistical uncertainty on the luminosity was only $\approx$0.01~dex, and we therefore reran the spectral fit using five other cloudy models: Exo-REM \citep{charnay2018}, petitCODE \citep{molliere2015}, DRIFT-PHOENIX \citep{helling2008}, BT-Settl \citep{allard2012}, and AMES-Dusty \citep{allard2001}. The dispersion on the retrieved luminosity, $\Delta\log{L/L_\odot} \approx 0.1$~dex, was adopted as the approximate systematic uncertainty. In the next section, we use the luminosity to quantify the mass of the planet.

\subsection{Bulk parameters and evolutionary constraints}
\label{sec:evolution_bulk}

In Sect.~\ref{sec:photometric_analysis} we showed that the $H$-band luminosity of HD\,135344\,Ab is consistent with a dusty mid L-type object with an approximate mass of $10$~$M_\mathrm{J}$. We now quantify the mass and other bulk parameters using the \texttt{species} toolkit. To do this, we fit the bolometric luminosity, $\log{L/L_\odot} = -3.9 \pm 0.1$, which we inferred from the atmospheric modeling, with a grid of evolutionary tracks that we interpolated as function of mass and age. The age of the system was applied as asymmetric normal prior by adopting the pre-main-sequence age of HD\,135344\,B, $11.9_{-5.8}^{+3.7}$~Myr \citep{garufi2018}. Similar to the spectral fit, a main limitation here are non-linear variations in the model grid, which can lead to inaccuracies and underestimated uncertainties. For the fit, we used the ATMO model \citep{phillips2020} because the cooling tracks seemed to be reasonably spaced for interpolation. The parameters were estimated with \texttt{MultiNest} \citep{feroz2008,buchner2014} by sampling the posterior distributions with 1000 live points.

\begin{figure}
\centering
\includegraphics[width=\linewidth]{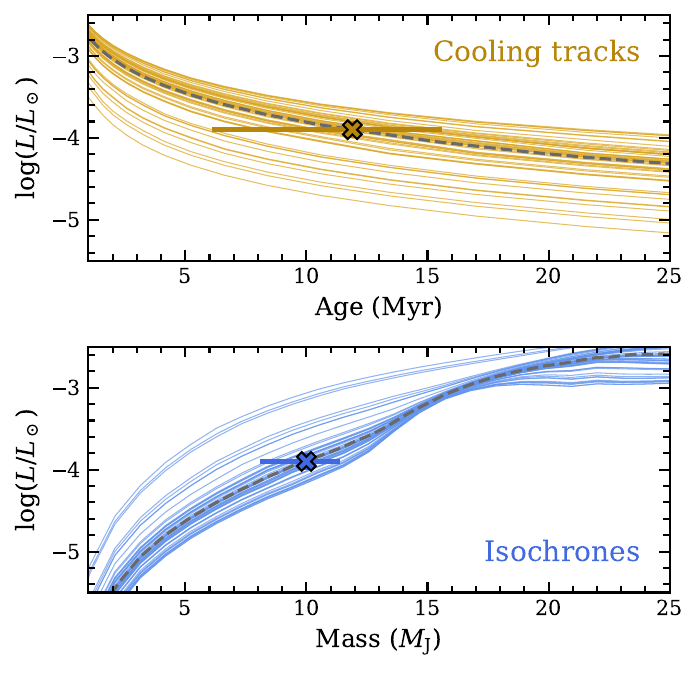}
\caption{Cooling tracks and isochrones inferred from the planetary luminosity. The colored lines show 50 random samples from the posterior distributions. The best-fit model is shown with the dashed line in each panel. The crosses represent the luminosity of HD\,135344\,Ab, $\log{L/L_\odot} = -3.9 \pm 0.1$, and the horizontal error bars indicate the prior age and posterior mass in the top and bottom panel, respectively.}
\label{fig:evolution}
\end{figure}

The mass and evolutionary constraints are presented in Fig.~\ref{fig:evolution}, which shows the isochrones and cooling tracks that describe the luminosity and age of HD\,135344\,Ab. We inferred a mass of $M_\mathrm{p} \approx 10$~$M_\mathrm{J}$, and the age is consistent with the prior, but is slightly more constrained (see Table~\ref{table:planet_param}). After the fit, we interpolated the evolutionary grid again for each mass-age sample in order to extract the related temperature, $T_\mathrm{eff} \approx 1585$, surface gravity, $\log{g} \approx 4.1$, and radius, $R \approx 1.5$. Table~\ref{table:planet_param} shows that the inferred bulk parameters are consistent within the considered credible regions with the parameter values estimated with the spectral modeling in Sect.~\ref{sec:atmospheric_modeling}. This suggests that the results are robust because we used two different, although correlated, approaches. The exception is the radius, which differs by 2$\sigma$ between the atmospheric and the evolution fit.

\section{Discussion and conclusions}
\label{sec:conclusions}

We have reported the direct discovery of a young giant planet at the A0V-type star HD\,135344\,A. The planet was detected through high-contrast imaging and interferometric observations. Our astrometric analysis of seven datasets, with a total baseline of four years, showed that the object is comoving with the central star. This was confirmed both by a common parallax and common proper motion. The inferred atmospheric and bulk parameters indicate a planetary nature, with a model-dependent mass of $M_\mathrm{p} \approx 10$~$M_\mathrm{J}$. The planet position changed by $\approx$30~mas and is slightly curved, which yielded a constraint on the orbital semi-major axis of approximately 15--20~au.

\begin{figure}
\centering
\includegraphics[width=\linewidth]{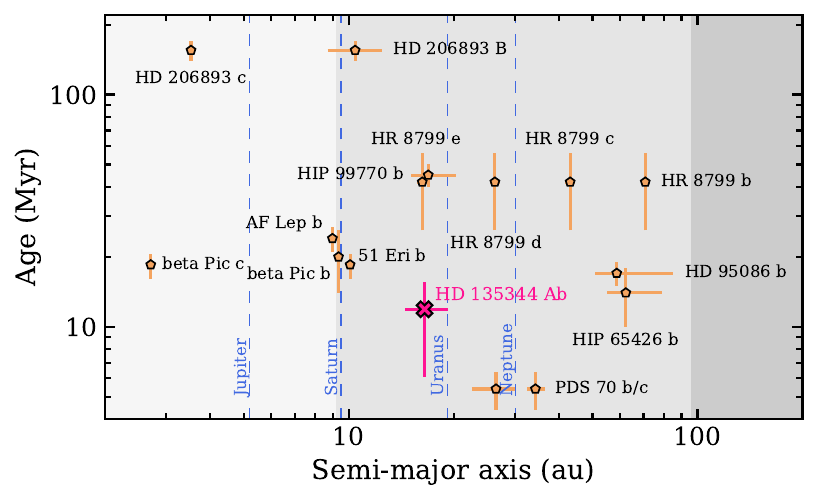}
\caption{Age vs. semi-major axis of directly imaged planets. We selected young companions with planetary masses (except for HD\,206893\,B), $M \lesssim 13$~$M_\mathrm{J}$, planet-to-star mass ratios of $q \lesssim \frac{1}{25}$, and orbits smaller than $\approx$100~au. The locations of the giant planets in the Solar System are indicated with vertically dashed lines, and the gray areas from left to right are separated by the approximate locations of the H$_2$O and CO$_2$ ice lines of an A0-type star \citep{oeberg2011}. The semi-major axes were retrieved from \texttt{whereistheplanet} when available \citep{whereistheplanet}, and from \citet{wang2021}, \citet{hinkley2023}, \citet{currie2023}, and \citet{derosa2023} otherwise. The ages were adopted from \citet{pecaut2012}, \citet{bell2015}, \citet{macintosh2015}, \citet{chauvin2017}, \citet{mueller2018}, \citet{garufi2018}, \citet{zuckerman2019}, \citet{miret-roig2020}, \citet{brandt2021}, and \citet{hinkley2023}. Systematic uncertainties on the ages (e.g., due to uncertain cluster membership) are not reflected by the error bars.}
\label{fig:age_separation}
\end{figure}

% Age-separation comparison, A/F stars, snowline
When we adopt the pre-main-sequence age of $\approx$12~Myr from the secondary star, HD\,135344\,Ab might be the youngest directly imaged planet that has fully formed and orbits on Solar System scales. Figure~\ref{fig:age_separation} shows a comparison with the ages and semi-major axes of other close-in directly imaged planets. Young directly imaged planets at small separations ($\lesssim$100~au) have only been detected at intermediate-mass stars. The planets in Fig.~\ref{fig:age_separation} all orbit A- and F-type stars, and the discovery of HD\,135344\,Ab at an A0-type star therefore follows that trend. This contrasts young planetary-mass objects on wide orbits ($\gtrsim$100~au), which are typically found at late-type stars \citep[e.g.,][]{bowler2014}. Because the orbit of HD\,135344\,Ab is relatively small, this planetary-mass companion is expected to have formed in a protoplanetary disk and is likely not the low-mass tail of binary star formation. The current planet location might be in the vicinity of the approximate snowline location for the spectral type of the host star (see Fig.~\ref{fig:age_separation}), although the planet may have migrated during its formation phase.

% Age discussion, impact on planet mass
In addition to the age constraint from HD\,135344\,B, \citet{ratzenbock2023b} recently used a clustering algorithm to map the star formation history of Sco-Cen. HD\,135344\,AB was associated with the $\phi$~Lup group in UCL, for which an isochrone age of $\approx$10~and $\approx$17~Myr was determined with two evolutionary models. An age of 10~Myr would match the pre-main-sequence age of HD\,135344\,B. Arguably, it is also be more consistent with the high IR excess of the secondary star, because it would be (even more) puzzling if a dust-rich disk were maintained up to 17~Myr. If the system were to be $\approx$17~Myr old, then the planet mass would still be in the planetary regime, $M_\mathrm{p} \approx 12$~$M_\mathrm{J}$, but close to the deuterium-burning limit.

% Disk depletion, formation history
In contrast to the secondary star, the circumstellar environment of HD\,135344\,A is already depleted, given the minor IR excess (see Appendix~\ref{app:stellar_parameters}). This is also consistent with a non-detection of a disk in scattered light in the SPHERE imagery. The origin of the different disk evolution timescales of the primary and secondary star is not known, but it might be related to a more efficient photoevaporation by the stronger radiation field of HD\,135344\,A. Disk lifetimes are indeed known to increase toward later spectral types \citep[e.g.,][]{luhman2022}. During the early evolution of the system, the primary star likely was a Herbig~Ae star with a protoplanetary disk in which HD\,135344\,Ab would have carved a wide gap during its formation. Dust- and gas-depleted cavities and gaps are quite common at Herbig Ae stars \citep[e.g.,][]{vandermarel2021}. These resolved substructures are signposts for forming planets, but only a few gap-carving planets have been detected (e.g., PDS\,70\,b/c; \citealt{keppler2018,haffert2019}). The discovery of close-in directly imaged planets such as HD\,135344\,Ab shows that Jovian planets might indeed have caused at least some of the large cavities. They might be more difficult to detect during formation because the dust has not yet been dispersed because even cavities are not fully cleared from small dust.

% Atmospheric constraints, spectral inference, bulk parameter accuracy
HD\,135344\,Ab is an appealing target for a spectral characterization with the next generation of large ground-based telescope facilities \citep[e.g.,][]{brandl2010} because of its small angular and physical separation from the star. The quality of the current measurements was sufficient to identify H$_2$O absorption in the low-resolution spectra and to infer the bulk parameters. A spectral inference of the molecular abundances will be more challenging because the planet is faint, but it might be feasible with KPIC, the fiber-fed high-resolution infrared spectrograph at Keck \citep{wang2024}, or with the enhanced sensitivity of the recent upgrade to GRAVITY+ \citep{gravity+2022}. Extending the SED from NIR to MIR wavelengths will increase the accuracy on the bolometric luminosity and other bulk parameters, and so will the extraction and calibration at short NIR wavelengths. Specifically, the fluxes at the blue end ($\lambda \lesssim 1.1$~$\mu$m) of the IFS spectrum are systematically higher than the model spectra in Fig.~\ref{fig:sed_planet}. We suspect that this is a bias in the spectral extraction, possibly due to the lower planet contrast, enhanced speckle noise, and/or reduced instrument transmission at the shortest wavelengths. Similar contaminating systematics are also seen at the short wavelengths in the IFS spectra of other faint planets \citep[e.g.][]{samland2017}. The effect did not impact the parameter estimation given the S/N of the spectra. Since the inferred luminosity is consistent with a planet mass of $M_\mathrm{p} \approx 10$~$M_\mathrm{J}$, but the $R_\mathrm{p}$ from the spectral fit is slightly too large given the mass and age constraint, this might imply that the $T_\mathrm{eff}$ and $R_\mathrm{p}$ inferred from the SED are somewhat under- and overestimated, respectively.

% Orbit constraints, periastron, misalignment with HD 135344 B disk
The orbital analysis yielded first constraints on the elements, given that the astrometry covers about 9\% of the orbital period, $P\approx 45$~yr. The semi-major axis has a precision of $\approx$2--3~au, but is correlated with the poorly constrained eccentricity, and therefore, also with the inclination. The posterior favors low to intermediate eccentricities, ruling out highly eccentric face-on orbits. However, this result requires confirmation through continued astrometric monitoring. The projected motion will be somewhat linear during the coming years, while the curvature will again increase toward periastron, which is in $2038.5 \pm 4.4$. Since the planet orbit has an high inclination, we conclude that the orbit is not coplanar with the protoplanetary disk of the secondary star in the HD\,135344\,AB binary system, as it is seen close to face-on ($i_\mathrm{disk} \approx 20$~deg; e.g. \citealt{perez2014}). This is also not surprising, given the large projected separation between the two stars ($\approx$2800~au); their circumstellar environments are therefore expected to have evolved independently.

% Dynamical mass, Gaia
Since the star was not observed by Hipparcos, we are unable to determine a proper motion anomaly. It will be interesting to analyze the astrometric measurements from Gaia DR4. Although the orbital period of HD\,135344\,Ab is $\approx$45~years and the baseline of DR4 is 5.5~years, an acceleration of the stellar proper motion might be detectable. When future Gaia epoch astrometry will be included in the orbit fit, a first constraint on the dynamical mass can be obtained, which will be valuable given the early evolutionary stage of the planet and its possible formation in a protoplanetary disk. Furthermore, a combined analysis of the absolute and relative astrometry could place constraints on the multiplicity of the planetary system because close-in directly imaged planets are often found in pairs (e.g., at $\beta$\,Pic, HD\,206893, and PDS\,70).

% Detection
The detection of HD\,135344\,Ab at only 3--4$\lambda/D$ demonstrates the powerful high-contrast and high-resolution capabilities of the SPHERE and GRAVITY instrument. This study also highlights the importance of high-precision astrometric measurements to fully disentangle orbital from background motion in a region of non-stationary background stars. A good portion of luck was involved with the discovery of HD\,135344\,Ab, however, because we caught the planet at a favorable separation along its inclined orbit. In the next 10 to 20 years, the angular separation with its star will decrease to $\approx$10--35~mas, which means that the planet would not have been discovered with SPHERE for a large fraction of its orbit.

% Outlook
Finally, direct imaging surveys have established that giant planets are rare at separations $\gtrsim$20~au \citep{nielsen2019,vigan2021}. The detection rate is expected to increase toward shorter separations, where radial velocity surveys have revealed a turnover point in the occurrence rates \citep{fernandes2019,fulton2021}. Gaia DR4 may reveal hints of similar close-in giant planets in star-forming regions, which will guide direct imaging searches and post-processing algorithms \citep[e.g.,][]{currie2023,winterhalder2024}. HD\,135344\,Ab might be part of a population of giant planets that could have formed in the vicinity of the snowline. These objects have remained challenging to detect since most surveys and observing strategies have not been optimized for such small separations.

\begin{acknowledgements}

Based on observations collected at the European Southern Observatory under ESO programmes 0103.C-0189(A), 105.20A8.001, 109.22ZA.004, 1104.C-0651(G), and 111.24FS.001. T.S.\ acknowledges the support from the Netherlands Organisation for Scientific Research (NWO) through grant VI.Veni.202.230. J.J.W.\ is supported by NASA XRP Grant 80NSSC23K0280. S.L.\ acknowledges the support of the French Agence Nationale de la Recherche (ANR), under grant ANR-21-CE31-0017 (project ExoVLTI). Part of this work was performed using the ALICE compute resources provided by Leiden University. This work used the Dutch national e-infrastructure with the support of the SURF Cooperative using grant no. EINF-1620. This research has made use of the Jean-Marie Mariotti Center \texttt{Aspro} service.

\end{acknowledgements}

\bibliographystyle{aa}
\bibliography{references}

\begin{appendix}

\section{Stellar parameters}
\label{app:stellar_parameters}

\begin{table*}
\caption{Stellar parameters of HD\,135344\,A.}
\label{table:stellar_param}
\centering
\bgroup
\def\arraystretch{1.25}
\begin{tabular}{L{2.0cm} C{2.2cm} C{1.5cm} C{4.5cm}}
\hline\hline
Parameter & Value & Units & Reference \\
\hline
RA (J2016) & +15 15 48.92 & hms & \citet{gaiadr3} \\
Dec (J2016) & -37 08 56.12 & dms & \citet{gaiadr3} \\
$\mu_\mathrm{RA}$ & $-18.74 \pm 0.05$ & mas yr$^{-1}$ & \citet{gaiadr3} \\
$\mu_\mathrm{Dec}$ & $-24.01 \pm 0.04$ & mas yr$^{-1}$ & \citet{gaiadr3} \\
$\varpi$ & $7.41 \pm 0.04$ & mas & \citet{gaiadr3} \\
\hline
TYCHO $B$ & $7.861 \pm 0.015$ & mag & \citet{hog2000} \\
TYCHO $V$ & $7.775 \pm 0.011$ & mag & \citet{hog2000} \\
\hline
Gaia $G$ & $7.7481 \pm 0.0028$ & mag & \citet{gaiadr3} \\
Gaia $G_\mathrm{BP}$ & $7.7687 \pm 0.0028$ & mag & \citet{gaiadr3} \\
Gaia $G_\mathrm{RP}$ & $7.6794 \pm 0.0038$ & mag & \citet{gaiadr3} \\
Gaia $G_\mathrm{RVS}$ & $7.6387 \pm 0.0051$ & mag & \citet{gaiadr3} \\
\hline
2MASS $J$ & $7.582 \pm 0.019$ & mag & \citet{cutri2003} \\
2MASS $H$ & $7.582 \pm 0.036$ & mag & \citet{cutri2003} \\
2MASS $K_\mathrm{s}$ & $7.563 \pm 0.023$ & mag & \citet{cutri2003} \\
\hline
WISE $W1$ & $7.538 \pm 0.028$ & mag & \citet{wright2010} \\
WISE $W2$ & $7.583 \pm 0.022$ & mag & \citet{wright2010} \\
WISE $W3$ & $7.140 \pm 0.017$ & mag & \citet{wright2010} \\
WISE $W4$ & $4.210 \pm 0.019$ & mag & \citet{wright2010} \\
\hline
SPHERE $H2$ & $7.59 \pm 0.03$ & mag & This work \\
SPHERE $H3$ & $7.58 \pm 0.03$ & mag & This work \\
SPHERE $K1$ & $7.57 \pm 0.02$ & mag & This work \\
SPHERE $K2$ & $7.56 \pm 0.02$ & mag & This work \\
\hline
SpT & A0V & & \citet{houk1982} \\
$T_\mathrm{eff}$ & $9540 \pm 100$ & K & This work \\
$\log\,g$ & $4.1 \pm 0.1$ & dex & This work \\
$\mathrm{[M/H]}$ & $\lesssim$0.05 & dex & This work \\
$R_\ast$ & $1.50 \pm 0.01$ & $R_\odot$ & This work \\
$A_V$ & $0.21 \pm 0.02$ & mag & This work \\
$\log{L/L_\odot}$ & $1.22 \pm 0.01$ & dex & This work \\
\hline
\end{tabular}
\egroup
\end{table*}

In this appendix, we analyze the spectral energy distribution (SED) of HD\,135344\,A. This is important for the calibration of the contrast measurements and, given the age of the system, to identify potential IR excess by circumstellar dust. Similar to fit of the near-infrared planet SED in Sect.~\ref{sec:atmospheric_modeling}, we used \texttt{species} \citep{stolker2020a} to model the stellar SED and retrieve the atmospheric parameters, in this case using the BT-NextGen model spectra \citep{allard2012}. The parameter posteriors were then used for computing synthetic photometry and spectra of the star, in order to convert the contrast to flux. The main stellar parameters are listed in Table~\ref{app:stellar_parameters}.

\begin{figure}
\centering
\includegraphics[width=\linewidth]{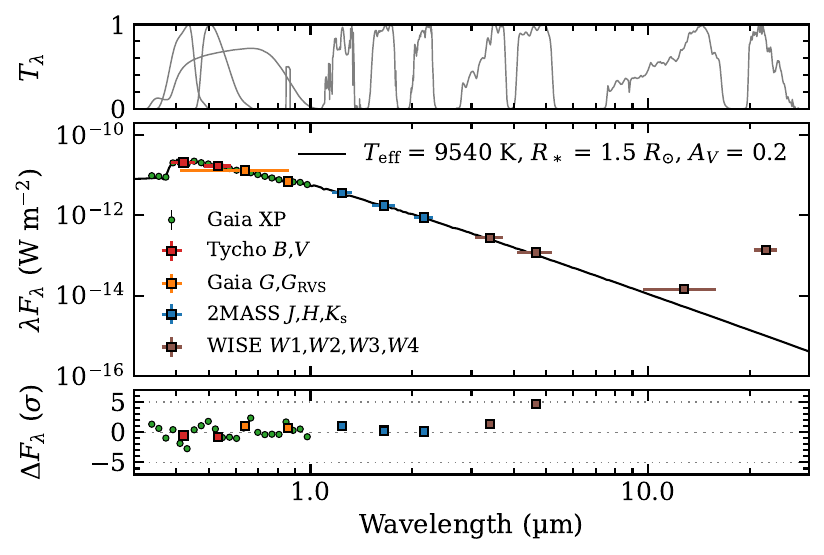}
\caption{Spectral energy distribution of HD\,135344\,A. The black line is the best-fit model spectrum and the colored markers are the photometric fluxes with horizontal error bars showing the FWHM of the filters. The top panel shows the filter profiles and the bottom panel the residuals relative to the measurement uncertainties. For clarity, every 15th wavelength of the Gaia XP spectrum is shown, whereas the full spectrum was used in the fit.}
\label{fig:sed_star}
\end{figure}

The parameter estimation is based on the low-resolution Gaia XP spectrum, and Gaia $G$ and $G_\mathrm{RVS}$, TYCHO $BV$, and 2MASS $JHK_\mathrm{s}$ photometry. We fitted an error bar inflation for the XP spectrum to account for the systematics that were seen as low-frequency oscillations. Similarly, we inflated the uncertainty of the $G_\mathrm{RVS}$ flux. The best-fit model spectrum has a goodness-of-fit of $\chi^2_\nu = 0.94$ and is compared with the data in Fig.~\ref{fig:sed_star}. The WISE fluxes were not included in the fit since the residuals revealed excess emission starting at WISE $W3$ ($\lambda_0 \approx 12$~$\mu$m) or perhaps already at $W2$ ($\lambda_0 \approx 4.6$~$\mu$m). The WISE photometry is however flagged as possibly contaminated by a diffraction spike of the secondary star, which has a high IR excess. Also, the initial WISE magnitudes are $\approx$0.2 and $\approx$2.0~mag fainter in $W3$ and $W4$, respectively, compared to the ALLWISE release \citep{wright2010}. Extracting robust photometry might be difficult since the two stars are hardly resolved at $W4$. So, while there seems to be evidence for IR excess, the magnitude is yet to determined. A more detailed analysis of the potential circumstellar disk will be deferred to a followup work.

\begin{figure}
\centering
\includegraphics[width=\linewidth]{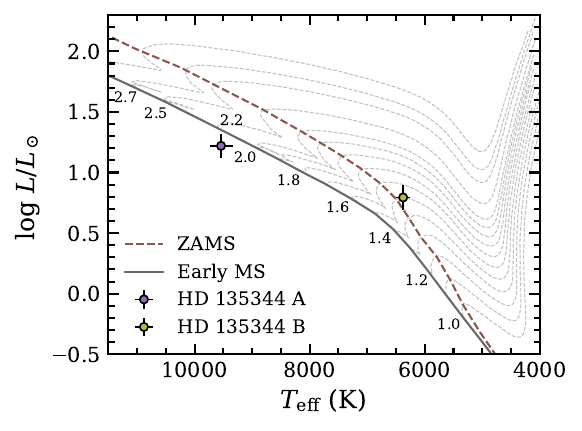}
\caption{Stellar evolutionary tracks in comparison with the HD\,135344\,AB binary system. The parameters of the primary star have been estimated in this work, but here showing inflated uncertainties of $\sigma_{T_\mathrm{eff}} = 200$~K and $\sigma_{\log{L/L_\odot}} = 0.1$~dex. The parameters of the secondary star have been adopted from \citet{fairlamb2015} and were corrected to the Gaia DR3 distance. The pre-main-sequence tracks are adopted from \citet{siess2000}. Stellar masses are provided next to the pre-main-sequence tracks in solar masses. The brown dashed line shows the zero age main-sequence (ZAMS; $L_\mathrm{nuclear} > 0.99 L_\mathrm{total}$), which is at 13.7~Myr for HD\,135344\,B ($M_\ast = 1.5$~$M_\odot$). The solid line is the early main-sequence, defined as the moment when the CNO cycle of intermediate-mass stars has reached its equilibrium.}
\label{fig:iso_contours}
\end{figure}

The retrieved stellar parameters are provided in Table~\ref{app:stellar_parameters}. The error bars reflect only the statistical uncertainties from the Bayesian inference, which might be be underestimated because model-dependent systematics are not accounted for. The effective temperature, $T_\mathrm{eff} \approx 9540$~K, is consistent with an A0V type star as evolutionary tracks predict about 9500~K for a stellar mass of $M_\ast = 2.2$~$M_\odot$ (see Fig.~\ref{fig:iso_contours}). The posterior distribution of the metallicity peaks near zero, the lower boundary of the model grid, indicating a preference for solar abundances. The visual extinction, $A_V \approx 0.2$, is consistent with the value derived for HD\,135344\,B, $A_V = 0.23 \pm 0.06$, by \citet{fairlamb2015}. The radius, $R_\ast \approx 1.5$~$R_\odot$, is smaller than the model prediction, $R_\ast = 1.8$~$R_\odot$. The radius acts as a flux scaling of the model spectrum, together with the normal prior for the parallax. The Gaia astrometric solution has a RUWE of 0.95 and the astrometric excess noise is 0.21~mas, indicating that the parallax measurement is sufficiently reliable to not bias the inferred stellar radius. From $T_\mathrm{eff}$ and $R_\ast$, we computed a bolometric luminosity of $\log L_\ast/L_\odot = 1.22 \pm 0.01$. The luminosity is low for an A0V type star (see Fig.~\ref{fig:iso_contours}), as a result of the small inferred stellar radius.

\end{appendix}

\end{document}